\pdfoutput=1
\documentclass[twocolumn,10pt]{article}
\usepackage[margin=0.5in]{geometry} 
\usepackage{amsmath}
\usepackage{amsfonts}
\usepackage[mathscr]{euscript}
\usepackage{dsfont}
\usepackage{amssymb}
\usepackage{graphicx}
\usepackage{url}
\usepackage{rotating}
\usepackage{verbatim}
\usepackage{color}
\usepackage{svg}
\usepackage{phonetic}
\usepackage{cite}
\usepackage{hyperref}
\usepackage[font=footnotesize,labelfont=bf]{caption}
\usepackage{bm}
\usepackage{upgreek}

\usepackage[runin]{abstract}
\DeclareFontFamily{OT1}{pzc}{}
\DeclareFontShape{OT1}{pzc}{m}{it}{<-> s * [1.10] pzcmi7t}{}
\DeclareMathAlphabet{\mathpzc}{OT1}{pzc}{m}{it}

\newcommand{\ii}{\bm{\mathsf{i}}}
\newcommand{\interior}{\mathbf{i}}
\newcommand{\exterior}{\mathbf{c}}
\newcommand{\powerset}{\mathcal{P}}
\newcommand{\rset}{\mathcal{S}}
\newcommand{\medialaxis}{\mathcal{M}}
\newcommand{\motion}{\mathcal{T}}
\newcommand{\R}{\mathpzc{R}}

\newcommand{\volume}{V}
\newcommand{\area}{A}

\newcommand{\com}[1]{}

\newcommand{\blue}[1]{\textcolor{blue}{#1}}

\title{Peg-in-Hole Revisited: A Generic Force Model for Haptic Assembly
    \footnote{This article was submitted on 12/02/2014 and published on 08/20/2015 in the ASME JCISE. For citation, please use:
    \protect\\
    \protect\\
        \blue{Behandish, Morad and Ilie\c{s}, Horea T., 2015. ``Peg-in-Hole Revisited: A Generic Force Model for Haptic Assembly.'' Journal of Computing and Information Science in Engineering, 15(4), p.041004.}
    \protect\\
    \protect\\
    A short version of this article was also presented in the ASME IDETC/CIE'2014 Conferences \cite{Behandish2014a}.
    }
}

\author{Morad Behandish and Horea T. Ilie\c{s}
\\
    {\small Departments of Mechanical Engineering and Computer Science and Engineering, University of Connecticut, USA}
}

\date{\small Technical Report No. CDL-TR-15-08, August 20, 2015}

\begin{document}

\maketitle

\noindent \hrule \vspace{5pt}
\begin{abstract}
The development of a generic and effective force model for semi-automatic or manual virtual assembly with haptic support is not a trivial task, especially when the assembly constraints involve complex features of arbitrary shape. The primary challenge lies in a proper formulation of the guidance forces and torques that effectively assist the user in the exploration of the virtual environment (VE), from repulsing collisions to attracting proper contact. The secondary difficulty is that of efficient implementation that maintains the standard 1 kHz haptic refresh rate. We propose a purely geometric model for an artificial energy field that favors spatial relations leading to proper assembly, differentiated to obtain forces and torques for general motions. The energy function is expressed in terms of a cross-correlation of shape-dependent affinity fields, precomputed offline separately for each object. We test the effectiveness of the method using familiar peg-in-hole examples. We show that the proposed technique unifies the two phases of free motion and precise insertion into a single interaction mode and provides a generic model to replace the {ad hoc} mating constraints or virtual fixtures, with no restrictive assumption on the types of the involved assembly features.
\end{abstract}

\vspace{5pt} \hrule \vspace{20pt}

\section{Introduction}

Computer haptics is an emerging technology in the modern virtual reality (VR) systems, with applications in areas as diverse as product design and prototyping, teleoperated and robot-assisted surgery, oral and dental implant operations, molecular simulation and training, rehabilitation systems, and gaming \cite{ElSaddik2011}.
The growth in the availability and popularity of this fairly recent technology imposes increasing demands for geometric modeling and computing algorithms, to deliver a realistic replication of the real-world experience in VEs as efficiently as possible. The efficiency issue appears more challenging in the case of haptic feedback, when compared to graphic rendering, due to the well-known physiological requirement of $1$ kHz refresh rate necessary for satisfactory tactile experience (especially, to acquire the necessary stiffness when manipulating rigid objects), while $30$$-$$60$ Hz is typically perceived as adequate for appealing to human vision \cite{ElSaddik2011}. 

One application domain that can tremendously benefit from an integration of multi-modal immersive user-interaction---i.e., an interaction through a multitude of human senses including vision, hearing, and more recently, touch---is computer-aided design and manufacturing (CAD/CAM).
Today, most engineering design tasks are heavily assisted by powerful and widely available computer simulation and visualization tools. Although a large subset of analysis and synthesis tasks have been partially (if not fully) automated, the designer's decision-making remains central to certain aspects of the design process. This in turn creates a demand for more effective human-computer interfaces to explore more efficient, creative, and cost-effective design solutions in semi-automatic setups. Haptic assistance has been found useful in several design activities that can benefit from domain expertise and cognitive capabilities of human operators (which are hard to formalize for full automation), such as conceptual design, design review and function validation, ergonomics and human factors evaluation, etc. \cite{Moreau2004,Bordegoni2006,Ferrise2010}.

Recently, an early-stage examination of different product life-cycle aspects related to design, manufacturing, maintenance, service, and recycling has been made possible by integrating VR tools into the modern CAD environments, a practice referred to as `virtual prototyping' \cite{Bullinger1999,Wang2002,Deviprasad2003}. Such an evaluation results in a significant reduction of time and cost associated with physical prototyping, and allows for the elimination of a large subset of design issues in the earlier stages of the process. Although they cannot yet completely replace physical prototypes, virtual prototypes are less expensive, more repeatable, and easily configurable for different variants, hence provide significant insight into the functionality of the product while eliminating redundant design trials and excessive tests \cite{Bordegoni2006}.
`Virtual assembly,' defined as a simulated assembly of the virtual representations of mechanical parts in an immersive 3D user interface using natural human motions, characterizes an important subset of virtual prototyping, to which applying haptic feedback has been shown particularly beneficial in terms of task efficiency and user satisfaction \cite{Gomes1999,Volkov2001}.

Computer-aided assembly planning (CAAP) typically deals with numerous part representations that are brought together by a set of pairwise `mating constraints.' In most commercially available digital environments such as the modern CAD software (e.g., CATIA, Pro/E, or NX) these constraints are classified into simple spatial relationships between the contact features, such as coplanarity of planar faces, coaxiallity of cylindrical features, and distance and angle offsets, and are {\it manually} specified by the designer. However, an {\it automatic} detection of these features on the one hand, and an identification of the correct one-to-one correspondence between them, on the other hand, for a given set of complex mechanical elements remain challenging in geometric modeling and design.
To the best of our knowledge, there is no universal model for automatic detection and matching of assembly features for objects of arbitrary shape, purely based on part geometry and not reliant on additional user input.

\subsection{Related Work} \label{sec_lit}

In the past two decades, there have been numerous studies focused on the development of immersive VEs for solving assembly and disassembly problems. These systems have used a variety of visualization tools (e.g., stereoscopic displays and goggles) and tracking devices (e.g., head tracking devices and data gloves) to assist the user in virtual object manipulation tasks.
More recently, an increasing number of studies have leveraged haptic devices to provide a more realistic assembly experience with force feedback, a thorough account of which would require a separate full paper. We refer the reader to \cite{Seth2011} for a more comprehensive survey of previous studies, and to \cite{Lim2010,Vance2011,Perret2013} for recent insight on current knowledge and expected future directions in haptic assembly. Here we provide a brief review of the most important techniques along with their advantages and limitations.

To realistically simulate the interactions in an assembly process, physically-based modeling (PBM) is used in most haptic-enabled assembly systems, where dynamic `part behavior' is simulated by integrating the (Newton+Euler's or Langrange's) equations of motion in real-time.
The most challenging set of computations are due to solving the `physical constraints' arising from contact between different objects in the scene, including rigid parts and subassemblies (typically imported from complex CAD models). There are two common approaches for computing the contact forces and torques in real-time.
The first method, referred to as the `penalty method,' employs simple force models that make {\it explicit} use of the collision response---e.g., a linear spring/damper model for computing the normal contact forces proportional to a measure of penetration between objects (or their offset shells) \cite{Hasegawa2003,Hasegawa2004} and a proper friction model using the normal pressure and the relative sliding/rolling kinematics to compute the tangential forces \cite{Mirtich1995,Hayward2000}.
This method is easy to implement and fast to integrate (given an efficient collision response and impact/friction modeling algorithm), but its robustness is heavily dependent on small integration time-steps to ensure minimal violations of constraints and rapid response to correct them.
The second method, referred to as the `constraint-based,' takes an {\it implicit} account of the unilateral contact constraints, and solves the more complex set of constrained equations of motion \cite{Renouf2005,Tching2008}.
It is more difficult to implement and takes more computing time, but it produces more accurate and reliable results and provides straightforward means to model tangential friction forces. Both methods are dependent on collision detection (CD), although they might use different CD information such as minimum distance, intersection volume, interpenetration depth, and contact normal vector.

There are several surveys of CD methods for rigid bodies \cite{Lin1998,Jimenez2001,Kockara2007} and flexible elements \cite{Teschner2005}.
Here, we restrict ourselves to review the most popular methods for real-time computations.
The classical polyhedral CD methods were used in the earliest systems for haptic assembly, such as Voronoi-clipping/marching methods (e.g., V-Clip \cite{Mirtich1998}, SWIFT \cite{Ehmann2000}, and SWIFT++ \cite{Ehmann2001} used in HIDRA \cite{Coutee2001,Coutee2002}), and oriented bounding box (OBB) tree-based methods (e.g, RAPID \cite{Gottschalk1996} used in MIVAS \cite{Wan2004}). However, these combinatorial techniques could not handle high-polygon models in haptic-enabled scenes due to the high frame rate requirement.
For a long time, uniform volumetric enumeration methods such as the Voxmap PointShell\textsuperscript{TM} (VPS) \cite{McNeely2005,McNeely2006} and its various improvements \cite{Barbic2007,Sagardia2008} became very popular for VR applications \cite{Renz2001,Johnson2001,Kim2003,Kim2004} and haptic assembly (e.g., used in the earliest versions of SHARP \cite{Seth2006,Seth2008}). VPS works by testing the moving objects represented by a shell of vertices and normals (i.e., the `pointshell'), against the stationary obstacles represented by a map of voxels (i.e., the `voxmap'). Although still being popular due to its simplicity and efficiency, the approximate representation scheme makes it ineffective for low-clearance assembly \cite{Seth2006,Seth2008}.

The PBM functionalities in virtual assembly applications (including collision response, impact/friction mechanics, and noon-smooth Lagrangian dynamics \cite{Tching2008}) are typically offered as part of a physics simulation engine (PSE).
Although a PSE+CD approach seems the most natural choice (at least in theory) for a virtual mimicry of real-world constrained motion, it is not effective in practice for final insertion in low-clearance assembly \cite{Seth2006,Seth2008}. This arises for at least two reasons: 1) numerical errors due to the approximate representations used in fast CD methods popular for haptic rendering; and 2) input noise due to hand vibrations and device errors. The former can be solved by using exact representations, which comes at the expense of computational burden. For example, later versions of SHARP \cite{Seth2007,Seth2010} employed the collision detection manager (CDM) module of D-Cubed, which makes direct use of Boundary Representation (B-rep) of CAD models. However, the latter difficulty persists, even with exact CD.

An alternative solution is to {\it artificially} introduce a set of bilateral (i.e., equality) constraints, rather than relying solely on groups of unilateral (i.e., inequality) constraints organically resulted from CD. The so-called constraint-based modeling (CBM) limits the number of degrees of freedom (DOF) of motion using `geometric constraints' (similar to mating constraints in CAD systems), and has been implemented using a variety of constraint management libraries \cite{Marcelino2003,Murray2004}.
One practical approach is to {\it manually} specify the mating constraints in close proximity of the final assembly configuration. The assembly constraints can be extracted from the CAD model or specified on-the-fly within the VE. For example, VADE \cite{Wang2003} and MIVAS \cite{Wan2004} directly imported pre-defined constraint information from Pro/E CAD models.
SHARP \cite{Seth2007,Seth2010} used the dimensional constraint manager (DCM) module of D-Cubed for defining and solving geometric constraints within the VE itself.
Rather than using the geometric semantics of the original parts, the virtual constraint guidance (VCG) method \cite{Tching2010a} relied on user-specified `virtual fixtures' \cite{Rosenberg1993}, which are added abstract and simple geometric elements rigidly attached to the fixed and moving parts---e.g., a pair of perpendicular planes intersecting at the axis of a cylindrical hole, to constrain and guide two points selected along the axis of a cylindrical peg.
A few recent studies attempted to {\it automatically} identify the assembly intent and associated geometric constraints by analyzing the semantic information of individual part geometries \cite{Iacob2008,Boussuge2012}, sometimes referred to as the automatic geometric constraints (AGC) method \cite{Seth2010}. This method relies on matching `functional surfaces' \cite{Iacob2011}---e.g, a cylindrical surface characterized by its axis and diameter, which could be used to predict the intended mating relation and associated trajectories when a peg is brought to the proximity of a hole. However, these methods are limited to matching simple (e.g., planar, cylindrical, spherical, and conical) geometric features. The effectiveness of both VCG and AGC methods relies heavily on either manual specification of the type of mating selected from a finite library of simple constraints, or heuristic models for identifying such mating pairs when corresponding simple geometric primitives are in proximity. A generic solution that automates the identification and pairing for features of arbitrarily complex surface geometry is missing.

Although it has been shown that limiting the motion DOF using geometric constraints supports highly accurate manipulation and positioning during low-clearance assembly in VEs \cite{Bowman2001}, the {ad hoc} nature of the constraint specification models and detection algorithms does not provide sufficient generality to completely replace generic physical constraints naturally imposed by CD. Consequently, the state-of-the-art in haptic assembly is based on a `two-phase' approach, i.e., to divide the process into a `free motion' phase accomplished with the help of CD engines, and a `fine insertion' phase using pre-specified or computer-predicted constraints \cite{Vance2011}.
There are two major difficulties faced in this approach. First, it requires developing mechanisms to detect the proximity to an insertion site, and to switch between the two modes. The implementations typically rely on CD between surface elements associated with the insertion constraints \cite{Seth2010} or between the user-defined virtual fixtures \cite{Tching2010a}. Once the alignment has been reached, part CD is switched off and the number of DOF is decreased to assist the user with the final insertion. Second, switching off part CD altogether results in a failure to detect a possible contact with geometry outside of the insertion area \cite{Perret2013}. To the best of our knowledge, the latter problem is also open.

\subsection{Contributions}

The current computational models for constraint-based assembly guidance are either 1) limited to the assembly of solids with very simple geometric features that are automatically detectable; or 2) heavily dependent on user input for constraint specifications. Both methods generally presume {a priori} knowledge of the type of contact surfaces that one deals with, and are not generalizable to support objects of arbitrary shape.
The majority of {ad hoc} solutions start from identifying the simplistic DOF-limiting constraints (e.g., restricting the motion to planar, cylindrical, spherical, or conical surfaces or their intersection curves), followed by what can be conceptualized as simple energy formulations to enforce those constraints (e.g., spring/damper models to penalize the violation of coplanarity or coaxiallity conditions).

We propose a generic and unified energy model for real-time assembly guidance that applies to objects of arbitrary shape. Our formulation starts from the part geometries and {\it directly} computes the guidance forces and torques from shape descriptors of interacting features. We do not make any simplifying assumption on the geometry of the mating features and show that implicit generalizations of the so-called virtual fixtures {\it automatically} appear in the form of a density distribution in the 3D space, called the Skeletal Density Function (SDF). The spatial overlapping of individual part SDFs generates an artificial potential energy (called the `geometric energy' field) which creates {\it attraction} forces and torques towards the proper alignment of assembly features. We show that the same energy model also provides {\it repulsion} forces and torques as a natural byproduct, in the case of collisions. Therefore, it unifies the two phases of free motion and precise insertion into a single interaction mode, thus avoids the duality and switch altogether.

\section{Formulation} \label{sec_form}

Given a set of mechanical components of a prospective assembly in a graphics- and haptics-enabled VE, the problem is to formulate a computational model to perform the following tasks:
\begin{enumerate}
    \item obtaining proper `shape descriptors' that capture the geometric and topological characteristics of the different components which are relevant to assembly and can be thought of as generic replacements for the {ad hoc} virtual fixtures;
    \item formulating a quantitative score function to measure the quality of the `geometric fit' between the shapes for arbitrary spatial configurations, based on overlapping the previously extracted shape descriptors;
    \item obtaining an artificial energy-field from the score model, whose gradient can be used as the guidance and constraint forces and torques during object manipulation in the VE (replacing the existing penalty methods based on linear spring/damper models).
\end{enumerate}
The goal is to develop a potent framework that performs these tasks without making any simplifying assumption on the shape, the intended function, or the proper spatial relationships of the parts.
The first step entails the most challenge from a theoretical point of view, since obtaining a quantitative description of the assembly features requires an understanding of the qualitative notion of a `proper fit,' and is not trivial for arbitrary geometry. The shape descriptors can be obtained in a preprocessing step for each rigid part. Therefore, the predominant computational challenges are pertaining to the real-time computations in the next two steps, due to the $1$ kHz haptic rendering rate requirement.
We examine each of these tasks in particular detail in the following sections, and introduce a novel paradigm that applies to arbitrarily complex shapes, without imposing any restricting assumption on the particular combination of contact features.

\subsection{Preliminaries} \label{sec_prem}

Following the good tradition of separating mathematical models \cite{Requicha1977a} from computational representations \cite{Requicha1980a}, we propose our formulation for general `solid' objects, defined as compact (i.e., bounded and closed), regular semi-analytic subsets of the Euclidean $3-$space $\rset \subset \powerset(\mathds{R}^3)$\footnote{The collection $\powerset(A) = \{ B ~|~ B \subset A \}$ denotes the `power set' of a set $A$, i.e., the set of all subsets of $A$.}
(i.e., `r-sets'). The regularity condition ensures that the set's `interior' $\interior S$, `exterior' $\exterior S$, and `boundary' $\partial S$ are well-defined notions, and prevents undesirable artifacts (such as `dangling' edges or isolated points that do not correspond to physically realizable shapes) \cite{Requicha1977a}. The semi-analytic requirement, on the other hand, guarantees triangulability of the set and prevents undesirable pathological behavior at the boundary \cite{Requicha1977a} and the skeleton \cite{Chazal2004}. Both conditions are sufficiently specific to enable theoretical developments as well as algorithmic tractability, yet general enough to encompass virtually all practically significant shapes for most solid modeling applications. We make an additional assumption that the boundary $\partial S$ is a piecewise $C^1-$manifold,
i.e., can be decomposed into a finite number of differentiable surface patches that are sewed together along sharp edges and corners. This enables formulating flux integrals over the boundary as a finite summation of surface integrals over those patches, each specified with well-defined normal vectors throughout their interiors.

It is worthwhile noting that our formulation does not impose, in principle, any restriction on the representation scheme, as long as it satisfies the informational completeness requirement \cite{Requicha1980a}---particularly, it suffices to support Euclidean distance queries and Point Membership Classification (PMC) tests \cite{Tilove1980a}. This applies to {\it exact} representations (e.g., parametric B-reps ranging from simple surfaces to nonuniform rational B-splines (NURBS) extracted from the CAD models) as well as {\it approximate} representations (e.g., triangular mesh or volumetric enumerations of the exported CAD models).  It is important to note that, especially when dealing with approximate representations, the employed shape descriptors must be stable and robust with respect to small perturbations in the boundary; otherwise they cannot be used effectively for designing computational algorithms.

\subsection{Shape Descriptors} \label{sec_des}

The assembly components need to be individually processed, each to be abstracted by certain shape descriptors that capture the most relevant geometric and topological characteristics to the virtual assembly task. This is probably the most challenging part of the entire process, especially when dealing with an infinitely large number of possibilities for complex surface features, each of which may or may not be the key determinant of proper assembly. The existing approaches to this and similar problems requiring feature recognition or characterization attempt to classify and match the potential contact features with respect to the common combinations of simple building blocks \cite{Iacob2008,Iacob2011,Boussuge2012}. These methods impose inevitable limitations on the geometric semantics of the objects. We take a completely different approach to avoid such limitations altogether.

\paragraph{Skeletal Density.}
The basic premise of our approach is that automatic identification of a proper fit in virtual assembly requires a quantification of the degree of effective geometric alignment, or {\it shape complementarity}, between pairs of objects. To achieve this, we make use of the new concept of continuous shape skeletons that we introduced in \cite{Behandish2014} for shape complementarity analysis of objects of arbitrary shape.

\begin{figure}
    \centering
    \includegraphics[width=0.48\textwidth]{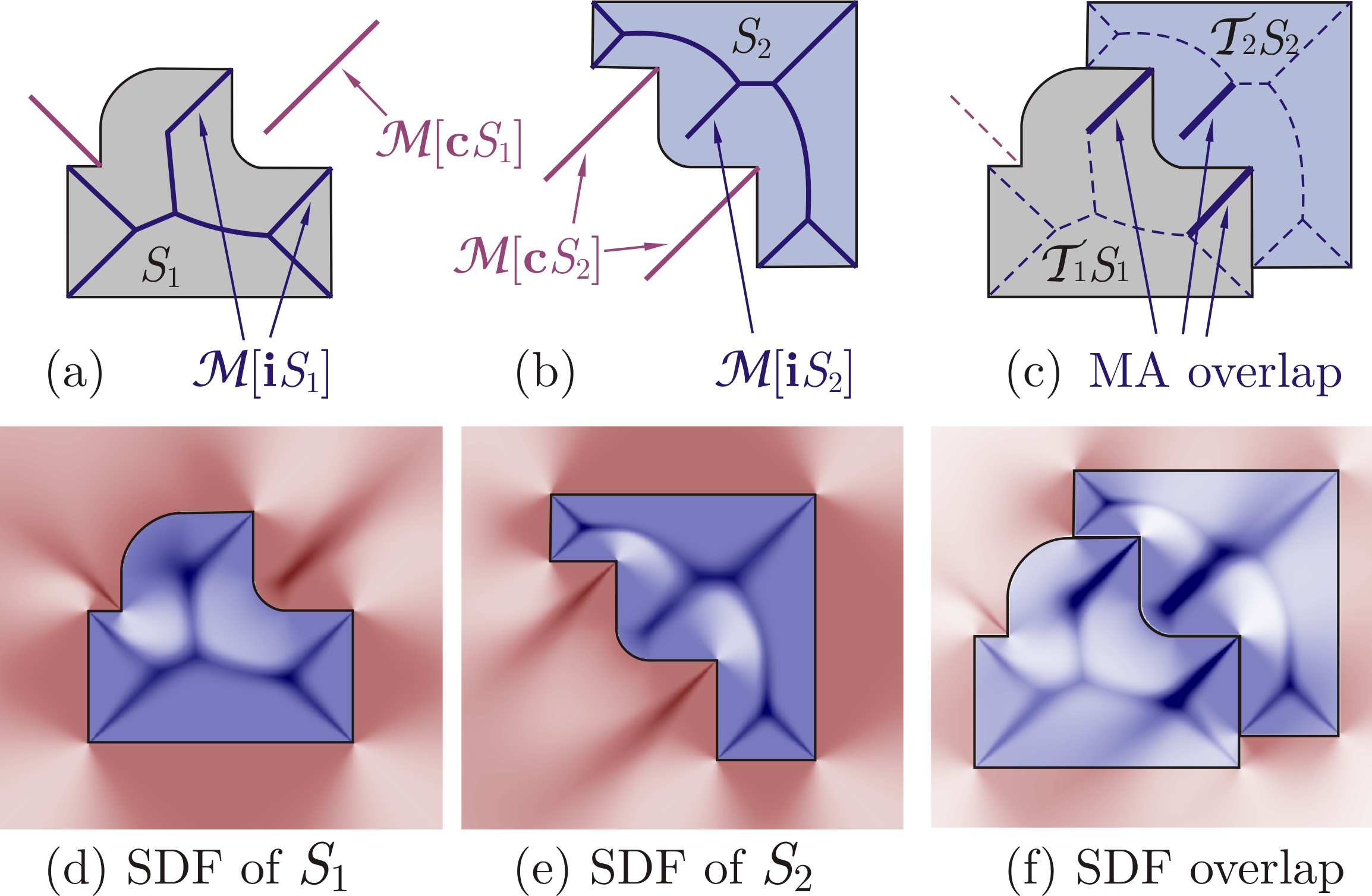}
    \caption{Assembly features captured by skeletal branches (a, b), which replace the virtual fixtures for assembly (c). The implicit skeletal density distribution (d, e) provides a robust substitute to facilitate measuring the overlap (f).} \label{figure1}
\end{figure}

Geometric skeletons, such as the medial axis (MA), can be regarded as abstractions of certain combinatorial, topological, and geometric information of the shape \cite{Lieutier2004}.
Figure \ref{figure1} (a, b) shows the MAs of interiors $\medialaxis[\interior S_{1,2}]$ and exteriors $\medialaxis[\exterior S_{1,2}]$ of the 2D r-sets $S_1, S_2 \in \rset$ (in this case $\rset \subset \powerset(\mathds{R}^2)$ for ease of illustration).
The MA branches can be used as abstractions of the shape for assembly features---e.g., the two branches associated with the sharp corners and the one branch associated with the fillet feature in Fig. \ref{figure1} (a, b). Therefore, one could try to overlap the external MA branches of one object with the internal MA branches of its mating object (and vice versa) to guide the assembly process, as in Fig. \ref{figure1} (c). This suggests using MA geometry as a generic replacement for the virtual fixtures \cite{Rosenberg1993} mentioned earlier. This treatment is applicable to features of arbitrarily complex shape, and requires no user specification prior to or during the assembly, hence liberates automatic computation of the guidance forces and torques regardless of the model complexity.

Unfortunately, the traditional definition of the MA is very unstable with respect to small perturbations in the boundary, making it extremely difficult to compute and prune \cite{Attali2009}. This motivated us to define a related concept in terms of a well-defined, space-continuous, and robust density distribution, called the Skeletal Density Function (SDF), whose sublevel sets in the limit are related to an implicit definition of the MA \cite{Behandish2014}.
Figure \ref{figure1} (d, e) shows the SDF field, that depends on a `thickness parameter' $\sigma > 0$. For $\sigma \ll 1$, the SDF value of the points on the MA (particularly those with more extensive nearest neighbors on $\partial S$) differentiates significantly from the points outside the MA. As $\sigma \rightarrow 0^+$, the SDF is related to the defining function of the MA under certain restricting conditions \cite{Behandish2014}.
We also showed that the SDF is a proper shape descriptor for the purpose of automatic prediction of assembly relations by means of comparative overlapping.
We briefly review the concepts that lead to the formal definition of SDF, skipping rigorous elaborations in favor of clarifying the main ideas. We do not intend to repeat the propositions, but rather to provide some insight into the applications of the SDF shape descriptors to define an energy model for haptic-assisted virtual assembly. We refer the reader to \cite{Behandish2014} for further mathematical details.

\paragraph{Distance Mapping.}
Given a solid $S \in \rset$ of arbitrary shape, we start by defining a Euclidean distance-based projection $\zeta: (\mathds{R}^3 \times \partial S) \rightarrow \mathds{C}$ of the boundary $\partial S$ to the complex plane, with respect to an arbitrary query point $\mathbf{p} \in \mathds{R}^3$ as
\begin{equation}
    \zeta(\mathbf{p}, \mathbf{q}; S) = \xi(\mathbf{p}; S) + \ii \eta(\mathbf{p}, \mathbf{q}), \label{eq_1}
\end{equation}
where the real-part $\xi(\mathbf{p}; S) = M(\mathbf{p}; S) \min_{\mathbf{q} \in \partial S} \| \mathbf{p} - \mathbf{q} \|_2$ is the signed Euclidean distance from the nearest neighbor on the boundary $\partial S$ to the query point $\mathbf{p} \in \mathds{R}^3$, whose sign is determined by the PMC function $M: \mathds{R}^3 \rightarrow \{-1, 0, +1\}$, i.e., $\xi < 0$ for interior points $(\mathbf{p} \in \interior S)$, $\xi = 0$ for boundary points $(\mathbf{p} \in \partial S)$, and $\xi > 0$ for exterior points $(\mathbf{p} \in \exterior S)$. The imaginary-part $\eta(\mathbf{p}, \mathbf{q}) = \| \mathbf{p} - \mathbf{q} \|_2$ is simply the $L_2-$distance between one particular boundary point $\mathbf{q} \in \partial S$ and the query point $\mathbf{p} \in \mathds{R}^3$.

The so-obtained $\zeta-$mapping can be conceptualized as a projection of the boundary $\partial S$ with respect to an arbitrary query point $\mathbf{p} \in \mathds{R}^3$, with the following properties:
\begin{enumerate}
    \item The real-part $\xi(\mathbf{p}, S)$ is constant for a fixed query point $\mathbf{p} \in \mathds{R}^3$, hence different boundary points $\mathbf{q} \in \partial S$ are mapped to a segment along the vertical line $\xi = $ const. on the complex plane, called the `complex spread' of the boundary and denoted as $\zeta(\mathbf{p}, \partial S; S)$.
    \item The location of the complex spread with respect to the imaginary axis is defined by the PMC; i.e., it is to the left, along, or to the right of $\xi = 0$, if the query point $\mathbf{p}$ is internal $(\mathbf{p} \in \interior S)$, on the boundary $(\mathbf{p} \in \partial S)$, or external $(\mathbf{p} \in \exterior S)$, respectively.
    \item By definition, if $|\xi| \leq \eta$ then $|\tan \angle \zeta| = \geq 1$, the equality being exclusive to the boundary points $\mathbf{q} \in \partial S$ that are the closest to the query point $\mathbf{p} \in \mathds{R}^3$ (i.e., $|\tan \angle \zeta| = 1$ iff $\mathbf{q}$ is the {\it exact} nearest neighbor of $\mathbf{p}$ on $\partial S$).
    \item For other boundary points at which $|\xi| < \eta$, the phase angle $\angle \zeta$ can be used as a determinant of the extent of normalized deviation for the boundary point $\mathbf{q} \in \partial S$ from being the nearest neighbor to the query point $\mathbf{p} \in \mathds{R}^3$; namely, $|\tan \angle \zeta| \leq (1+\epsilon)$ identifies the $\epsilon-$approximate nearest neighbors ($\epsilon-$ANNs).
\end{enumerate}

The MAs of interior $\medialaxis[\interior S]$ and exterior $\medialaxis[\exterior S]$ of an r-set $S \in \rset$ are defined as the loci of points in the interior $\mathbf{p} \in \interior S$, and exterior $\mathbf{p} \in \exterior S$, respectively, that have strictly more than one {\it exact} nearest neighbor on the boundary \cite{Attali2009},\footnote{Although we loosely refer to the MA as a type of shape skeleton, the exact definitions of `medial axis,' `skeleton,' and `cut locus' (i.e., the closure of MA) are different (but closely related) for general open sets \cite{Lieutier2004,Chazal2004}.}
which can be implicitly defined by counting the number of points on the boundary $\partial S$ that map to the same complex point $\zeta \in \mathds{C}$ with $|\tan \angle \zeta| = 1$.
The strict condition on the existence of at least two points $\mathbf{q}_1, \mathbf{q}_2 \in \partial S$ that {\it exactly} satisfy $\eta(\mathbf{p}, \mathbf{q}_1) = \eta(\mathbf{p}, \mathbf{q}_2) = |\xi(\mathbf{p}; S)|$ makes the MA extremely unstable with respect to $C^0-$ and $C^1-$perturbations of the boundary resulting from noise/errors in shape data, since a small perturbation of the surface geometry may result in large changes in the topology and geometry of the MA \cite{Chazal2004,Attali2009}. In addition to the extremely difficult computation and refinement of the MA in practice, another challenge is of obtaining a shape complementarity score function that changes continuously with deviations in spatial relationships---i.e., a score function that properly rewards approximate overlap between MA branches, and penalizes separation between them. This is particularly difficult because not all branches are supposed to overlap (see Fig. \ref{figure1} (c)) and those branches that do overlap might not exactly be coincident, especially in the presence of approximations. We solved these problems in \cite{Behandish2014} by relaxing the aforementioned strict condition using {\it approximate} nearest neighbors, and by redefining the skeletal shape descriptors as space-continuous scalar fields whose overlaps can be quantified easily and robustly by SDF inner products (i.e., function integrals).

\paragraph{Complex Kernel.}
For a given query point $\mathbf{p} \in \mathds{R}^3$, rather than counting the number of boundary points $\mathbf{q} \in \partial S$ for which $|\tan \angle \zeta| = 1$, which gives a discontinuous integer-valued defining function of the MA, we define a space-continuous complex-valued density function called the Skeletal Density Function (SDF). This is realized by first defining a kernel $\phi_\sigma: (\mathds{C} - \{0\}) \rightarrow \mathds{C}$ that takes the deviation of $|\tan \angle \zeta|$ from unity for the points on the complex spread of the boundary, and assigns a larger density to the query points $\mathbf{p} \in \mathds{R}^3$ that have more extensive $\epsilon-$ANNs. This means a denser patch of points on $\zeta(\mathbf{p}, \partial S; S)$ with $|\tan \angle \zeta| \leq (1+\epsilon)$. The following particular definition serves this purpose:
\begin{equation}
    \phi_\sigma(\zeta) = \frac{\lambda(\zeta)}{\sqrt{2\pi}} \frac{1}{\zeta^2} g_\sigma \left(|\tan \angle \zeta| - 1 \right), \label{eq_2}
\end{equation}
where $\lambda(\zeta) = +\lambda_1$ if $\mathrm{Re}\{\zeta\} > 0$, and $\lambda(\zeta) = -\lambda_2$ if $\mathrm{Re}\{\zeta\} < 0$, using $0 < \lambda_1 < \lambda_2$ for reasons to become clear in Section \ref{sec_comp}. This gives a different sign and weight to the $\phi-$kernel based on whether $\mathbf{p}$ is external ($\xi > 0$) or internal ($\xi < 0$) to $S$, respectively. $g_\sigma(x) = (\sqrt{2\pi} \sigma)^{-1} e^{-\frac{1}{2}(x/\sigma)^2}$ is the Gaussian function with thickness factor of $\sigma > 0$, which is meant to serve as the `medial' component of the $\phi-$kernel, namely, to assign larger densities to the medial points with $|\tan \angle \zeta| \approx 1$ with a continuous decay that is controllable by the parameter $\sigma$. The `proximal' component $(\sqrt{2\pi}\zeta^2)^{-1}$, on the other hand, is provided to enforce two effects, namely, 1) an inverse-square decay of the skeletal density when the query point moves away from the boundary; and 2) a phase difference of $\angle \phi = -2\angle \zeta$ which results in $\angle \phi \approx \pi \mp \pi/2$ for the high-density medial points with $\angle \zeta \approx \pi/2 \pm \pi/4$ (i.e., $|\tan \zeta| \approx 1$). In the context of haptic-assisted assembly, we will see that the former translates into a `gravity' force between the assembly components (i.e., favoring proximity, hence the name of the term), while the latter induces a sign changing mechanism that underlies the switch between the attraction and repulsion modes, when the parts are about to reach proper contact versus when they are about to penetrate, respectively.

\paragraph{Affinity Function.}
The next step is to apply the custom $\phi-$kernel to the complex spread of the object under consideration to obtain the SDF, also known in this context as the `affinity function' $\rho_\sigma: (\mathds{R}^3 - \partial S) \rightarrow \mathds{C}$:
\begin{equation}
    \rho_\sigma(\mathbf{p}; S) = \oint_{\partial S} \phi_\sigma \Big[ \zeta\left(\mathbf{p}, \mathbf{q}; S\right) \Big] ~ d\area_{\bot}, \label{eq_3}
\end{equation}
where $d\area_{\bot}$ is the area element normal to the line that connects $\mathbf{p} \in \mathds{R}^3$ to $\mathbf{q} \in \partial S$; that is, the infinitesimal area of the projection of the surface element $d\area$ at the boundary point $\mathbf{q}$ on a sphere centered at the query point $\mathbf{p}$ with a radius of $\eta(\mathbf{p}, \mathbf{q})$.
If we let $\mathbf{v} = (\mathbf{p} - \mathbf{q})/\eta(\mathbf{p}, \mathbf{q})$ be the unit `gaze vector,' then we obtain $d\area_{\bot} = (\mathbf{v} \cdot \mathbf{n}) d\area$ and (\ref{eq_3}) becomes a flux integral of the radial vector field $\phi_\sigma(\zeta) \mathbf{v} : (\mathds{C} - \{ 0 \}) \rightarrow \mathds{C}^3$ over the oriented piecewise $C^1-$manifold $\partial S$.
Substituting for $\phi(\zeta)$ from (\ref{eq_2}), and letting $1 + \epsilon = |\tan \angle \zeta| = \eta / |\xi|$, the following results from (\ref{eq_3}):
\begin{equation}
    |\rho_\sigma(\mathbf{p}; S)| = \oint_{\partial S} \frac{ \frac{|\lambda|}{\sigma} e^{-\frac{1}{2} (\epsilon/\sigma)^2}}{1 + (1 + \epsilon)^{-2}} ~ \frac{d\area_{\bot}}{2\pi\eta^2}, \label{eq_4}
\end{equation}

We showed in \cite{Behandish2014} that the exact SDF given in (\ref{eq_3}) can be approximated with a {\it truncated} SDF that is integrated over the regions of the boundary that form the $\epsilon-$ANNs of $\mathbf{p}$, i.e., the patches of the boundary that lie within a closed spherical shell of internal radius $|\xi|$ and external radius $(1+\epsilon)|\xi|$, and proved truncation error bounds on the approximation. This can be explained in simple terms by the fact that the Gaussian term in (\ref{eq_2}) decays exponentially for the points outside the $\epsilon-$approximate nearest neighborhood; in fact, for $|\tan \angle \zeta| > (1 + \epsilon)$, the exponential term is at most $e^{-\frac{1}{2}(\epsilon/\sigma)^2}$, which is in turn less than $10^{-4}$ if we choose $\epsilon > 3.4 \sigma$.

It is also interesting to note that $d\gamma = d\area_{\bot}/(4\pi\eta^2)$ is the infinitesimal solid angle by which the query point $\mathbf{p} \in \mathds{R}^3$ observes the surface element $d\area$ at $\mathbf{q} \in \partial S$. Therefore, the surface integral in (\ref{eq_4}) aggregates the $\phi-$kernel over the $\epsilon-$ANNs to the query point on the boundary, assigning weights proportional to the spatial angles by which they are observed. This explains the choice of the inverse-square law in the proximal term for the $\phi-$kernel in (\ref{eq_2}) over any other possible decay function.

\begin{figure}
    \centering
    \includegraphics[width=0.48\textwidth]{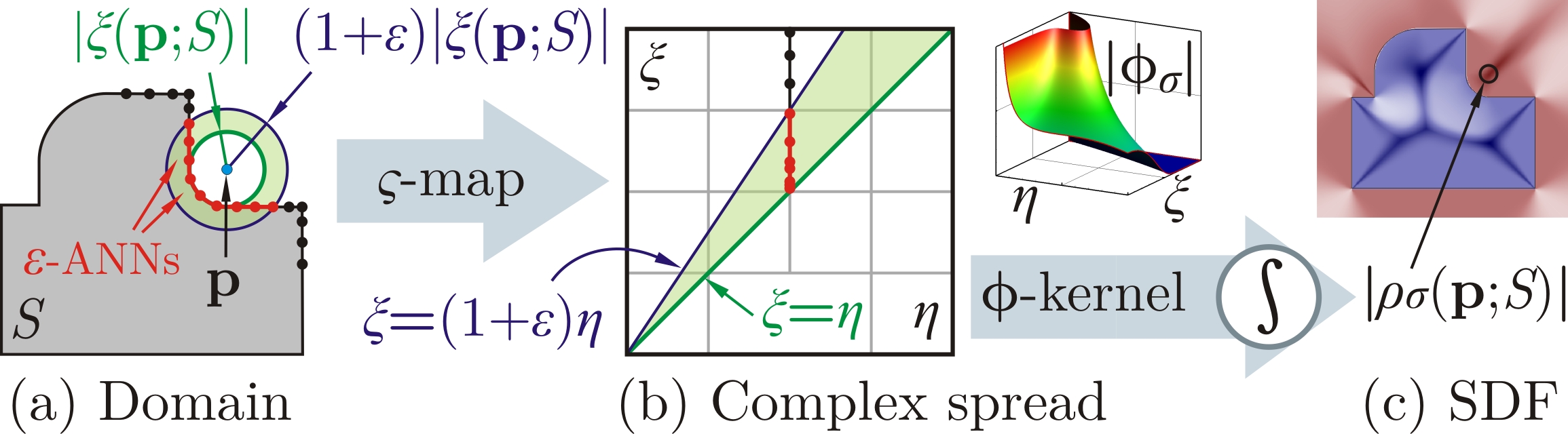}
    \caption{The affinity computation is decomposed into two steps: a $\zeta-$projection in (\ref{eq_1}) that characterizes the distance distribution as observed from the query point $\mathbf{p} \in \mathds{R}^3$, followed by applying the $\phi-$kernel in (\ref{eq_2}).} \label{figure2}
\end{figure}

Figure \ref{figure2} provides a schematic description of the SDF computation process, decomposed in principle into 1) computing a different representation of the object based on the Euclidean distance geometry, i.e., the $\zeta-$projection of the boundary with respect to different query points; and 2) the application of a custom $\phi-$kernel to determine the distance distribution criteria, based on which the skeletal density is assigned in the $3-$space. See \cite{Behandish2014} for examples of using different kernels and their applications.

\paragraph{Affinity Gradient.}
The gradient of the affinity function $\nabla \rho_\sigma = d \rho_\sigma / d \mathbf{p}: \mathds{R}^3 \rightarrow \mathds{C}^3$ (to be used in Section \ref{sec_comp} to compute the guidance forces and torques) can be computed from (\ref{eq_3}) by applying the chain rule for differentiation:

\begin{equation}
    \nabla \rho_\sigma(\mathbf{p}; S) = \oint_{\partial S} \left[ \frac{\partial \phi_\sigma}{\partial \xi}(\zeta) \nabla \xi + \frac{\partial \phi_\sigma}{\partial \eta}(\zeta) \mathbf{v} \right] ~ d\area_{\bot}, \label{eq_3a}
\end{equation}
where $\nabla \xi = d \xi / d \mathbf{p}: \mathds{R}^3 \rightarrow \mathds{R}^3$ is the {\it extended} gradient of the signed Euclidean distance function \cite{Lieutier2004},
and the partial kernel derivatives $\partial \phi_\sigma /\partial \xi, \partial \phi_\sigma /\partial \eta : \mathds{C} \rightarrow \mathds{C}$ can be obtained directly from (\ref{eq_2}).

\subsection{Shape Complementarity} \label{sec_comp}

In a complex virtual assembly scene with many components, the analysis of the proper contact between all parts can be broken down in a bottom-up fashion into pairwise matching between the parts, and then between the resulted subassemblies, with an incremental growth of the number of constituents. This view is compatible with the actual process of semi-automatic haptic-enabled assembly, when the user drags and places the parts and resulting subassemblies one at a time.

\paragraph{Score Function.}
For a pair of solids $S_1, S_2 \in \rset$ (each solid representing one part or subassembly as a single object), the motion of both objects at any instant of time can be described by the transformations $\motion_1, \motion_2 \in \mathrm{SE}(3)$ that relate the absolute coordinate frame to an orthonormal triad attached to each object; $\mathrm{SE}(3) \cong \mathrm{SO}(3) \ltimes \mathrm{T}(3)$ represents the special Euclidean group, i.e., combination of proper orthogonal rotations $\mathrm{SO}(3)$, and translations $\mathrm{T}(3)$, together representing all possible rigid body motions. We proposed in \cite{Behandish2014} that the shape complementarity score for this configuration can be obtained as a cross-correlation of individual SDF fields over the $3-$space:
\begin{equation}
    f_\mathrm{SC}\big( \motion_{1,2}; S_{1,2} \big) = \int_{\mathds{R}^3} \rho_\sigma \big(\mathbf{p}'; \motion_1 S_1 \big) \rho_\sigma \big( \mathbf{p}'; \motion_2 S_2 \big) d\volume, \label{eq_5_0}
\end{equation}
where $\rho_\sigma(\mathbf{p}'; S)$ is substituted from (\ref{eq_3}); the integration variable is $\mathbf{p}' = (x_1, x_2, x_3) \in \mathds{R}^3$ and $d\volume = dx_1 dx_2 dx_3$ is the volume element.
The formulation can be significantly simplified using a kinematic inversion. Let $\mathbf{p} = \motion_1^{-1} \mathbf{p}'$ be the new coordinates of the query point measured with respect to a frame attached to $S_1$, and $\motion = \motion_1^{-1}\motion_2 \in \mathrm{SE}(3)$ be the motion of $S_2$ observed from that same frame. Noting that the SDF is solely formulated based on distance distributions, it is invariant under rigid body transformation, i.e., the affinity field moves rigidly with the object, hence $\rho_\sigma(\mathbf{p}; \motion S) = \rho_\sigma(\motion^{-1}\mathbf{p}; S)$ for all $S \in \rset$ and $\motion \in \mathrm{SE}(3)$. Using this property and rearranging the terms in (\ref{eq_5_0}) give
\begin{equation}
    f_\mathrm{SC}\big( \motion; S_1, S_2 \big) = \int_{\mathds{R}^3} \rho_\sigma \big(\mathbf{p}; S_1 \big) \rho_\sigma \big(\motion^{-1} \mathbf{p}; S_2 \big) d\volume. \label{eq_5}
\end{equation}
In practice, (\ref{eq_5}) is evaluated over a bounded cubic region of $\mathds{R}^3$ that is large enough to cover the high density segments of the SDF branches, noting the inverse-square decay of $\rho_\sigma(\mathbf{p}; S)$ in (\ref{eq_3}).
The integral in (\ref{eq_5}) can be interpreted as an assessment of the degree of overlap between the continuous internal and external skeletons of one object, and the internal and external skeletons of the other object, hence four possible combinations contributing differently to the overall score function. We describe the four scenarios in simple terms to convey an intuitive understanding of (\ref{eq_5_0}) and (\ref{eq_5}). Consider a volume element at a point that belongs to a high skeletal density region of both $\motion_1 S_1$ and $\motion_2 S_2$, hence contributing significantly to the integral. Assume that the distance distribution over the $\epsilon-$ANNs on the boundaries of the two objects as observed from the query point are similar, hence the medial and proximal components of the SDF are equally high with respect to both shapes, making the $\lambda-$function in (\ref{eq_2}) decide the separations between the following cases:
\begin{itemize}
    \item[\textbullet] if the point is external to object-1 and internal to object-2, i.e., $\mathbf{p}' \in \left[ \exterior(\motion_1 S_1) \cap \interior(\motion_2 S_2) \right]$ thus $\mathbf{p} \in \left[ \exterior S_1 \cap \interior(\motion S_2) \right]$ then $\rho_1 \rho_2 \propto (- \ii \lambda_1) (+ \ii \lambda_2) = + \lambda_1 \lambda_2 > 0$;
    \item[\textbullet] if the point is internal to object-1 and external to object-2, i.e., $\mathbf{p}' \in \left[ \interior(\motion_1 S_1) \cap \exterior(\motion_2 S_2) \right]$ thus $\mathbf{p} \in \left[ \interior S_1 \cap \exterior(\motion S_2) \right]$ then $\rho_1 \rho_2 \propto (+ \ii \lambda_2) (- \ii \lambda_1) = + \lambda_2 \lambda_1 > 0$;
    \item[\textbullet] if the point is internal to object-1 and internal to object-2, i.e., $\mathbf{p}' \in \left[ \interior(\motion_1 S_1) \cap \interior(\motion_2 S_2) \right]$ thus $\mathbf{p} \in \left[ \interior S_1 \cap \interior(\motion S_2) \right]$ then $\rho_1 \rho_2 \propto (+ \ii \lambda_2) (+ \ii \lambda_2) = - \lambda_2^2 < 0$;
    \item[\textbullet] if the point is external to object-1 and external to object-2, i.e., $\mathbf{p}' \in \left[ \exterior(\motion_1 S_1) \cap \exterior(\motion_2 S_2) \right]$ thus $\mathbf{p} \in \left[ \exterior S_1 \cap \exterior(\motion S_2) \right]$ then $\rho_1 \rho_2 \propto (- \ii \lambda_1) (- \ii \lambda_1) = - \lambda_1^2 < 0$;
\end{itemize}
\begin{figure}
    \centering
    \includegraphics[width=0.48\textwidth]{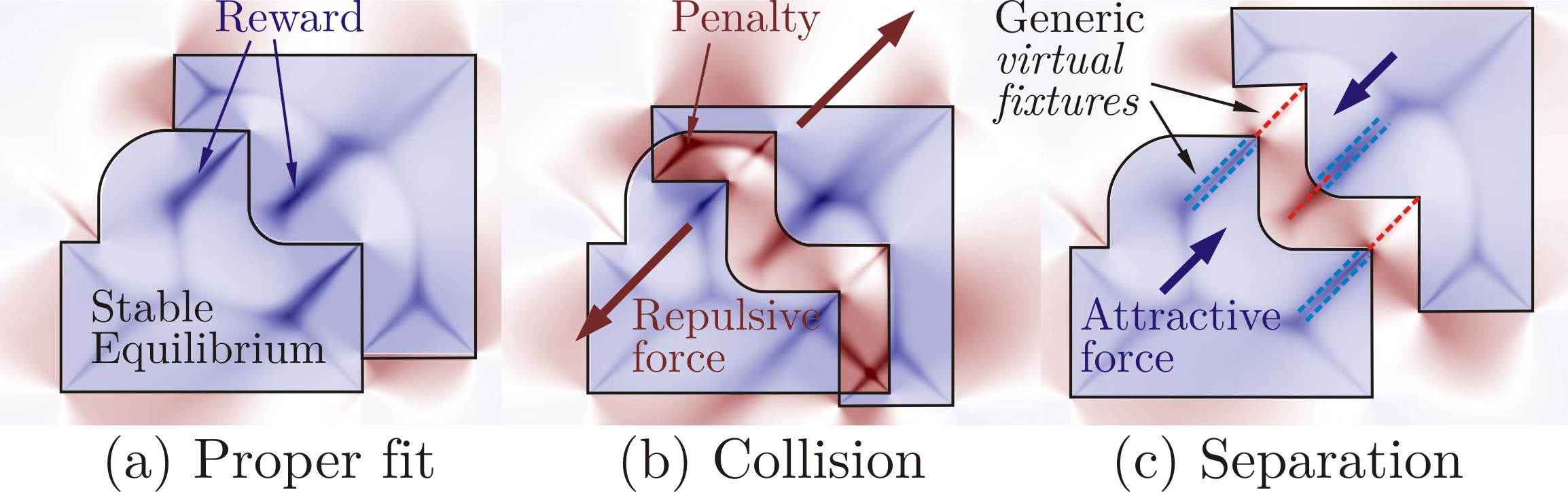}
    \caption{Possible spatial relations and the corresponding interactions. The generic virtual fixtures practically restrict the DOF if the stiffness properties (i.e., 2nd-order partial derivatives of $E_\mathrm{G}$ at the energy well) are large enough.} \label{figure3}
\end{figure}

The first two cases characterize the `proper fit' alignment between the two objects, since the exterior of one object is aligned with the interior of another with similar distance geometries, carrying the hint of a proper complementary feature (Fig. \ref{figure3} (a)). On the other hand, the third case implies `collision,' since the interior points are being overlapped, and should be strictly prohibited (Fig. \ref{figure3} (b)). Lastly, the fourth case suggests a `separation' at the observation point, which amounts less to a conclusion about the quality of fit (Fig. \ref{figure3} (c)). Hence if we choose $\lambda_1 = O(1)$ and $\mathfrak{p} := \lambda_2/ \lambda_1 \gg 1$, then the first two cases contribute a positive-real {\it reward} of $\propto O(\mathfrak{p})$ to the score function, the third term contributes a large negative-real {\it penalty} of $\propto O(\mathfrak{p}^2)$, and the last term contributes a smaller penalty of $\propto O(1)$. The ratio $\mathfrak{p}$ is thus called the `penalty factor.' These of course describe the distance geometry as observed from a single query point under consideration, which is why the overlap is integrated over different observation points via (\ref{eq_5}) to obtain the cumulative effect.

\paragraph{Motion Decomposition.}
To simplify the subsequent development, let us decompose the motion into the translational component $\mathbf{t} \in \mathrm{T}(3)$ described by a $3-$tuple $(\mathpzc{t}_1, \mathpzc{t}_2, \mathpzc{t}_3) \in \mathds{R}^3$, and the rotational component $\mathpzc{R} \in \mathrm{SO}(3)$ represented by a $3\times3$ proper orthogonal matrix $[\R]_{3\times3}$. As a result of the definition, the transformation sequence applies as $\motion \mathbf{p} = (\R \mathbf{p}) + \mathbf{t}$ hence $\motion^{-1} \mathbf{p} = \R^\mathrm{T} (\mathbf{p} - \mathbf{t})$.\footnote{Note that $\R^{-1} = \R^\mathrm{T}$ for all $\R \in \mathrm{SO}(3)$, i.e., the inverse of a proper rotation with $\mathrm{det}(\R) = +1$ is the same as its transpose.}
Substituting for the latter in (\ref{eq_5})
\begin{equation}
    f_\mathrm{SC} \big( (\R,\mathbf{t}); S_1,S_2 \big) = \int_{\mathds{R}^3} \rho_1(\mathbf{p}) \Big[ \rho_2 \circ \R^\mathrm{T} (\mathbf{p} - \mathbf{t}) \Big] d\volume. \label{eq_5a}
\end{equation}
where the functions $\rho_{1,2}(\mathbf{p}) = \rho_\sigma(\mathbf{p}; S_{1,2})$ are independent of the motion parameters, and the inverse rotation operator is treated as a function $\R^\mathrm{T} : \mathds{R}^3 \rightarrow \mathds{R}^3$.

\paragraph{Score Gradient.}
For a function $f_\mathrm{SC} : \mathrm{SE}(3) \rightarrow \mathds{C}$ whose domain is not a vector space, the generalized gradient function $\nabla f_\mathrm{SC} = (d f_\mathrm{SC} / d\R, d f_\mathrm{SC} / d\mathbf{t}) : \mathrm{SE}(3) \rightarrow \mathds{C}^6$ is composed of a 3D translational and a 3D rotational gradient vectors, characterizing the rate of change of the function with respect to infinitesimal translations and rotations, respectively.

The translational gradient function $df_\mathrm{SC}/d\mathbf{t} : \mathrm{SE}(3) \rightarrow \mathds{C}^3$ is computed using basic concepts from linear algebra, since the translation space $\mathrm{T}(3) \cong \mathds{R}^3$ is a vector space. Differentiating (\ref{eq_5a}) and using the chain rule we obtain
\begin{equation}
    \langle \frac{d f_\mathrm{SC}}{d \mathbf{t}}, \mathbf{e} \rangle = - \int_{\mathds{R}^3} \rho_1(\mathbf{p}) \Big[ \nabla \rho_2 \circ \R^\mathrm{T} (\mathbf{p} - \mathbf{t}) \Big] \cdot (\R^\mathrm{T} \mathbf{e}) d\volume, \label{eq_5b}
\end{equation}
where $\mathbf{e} \in \mathds{R}^3$ represents any direction in the vector space $\mathrm{T}(3) \cong \mathds{R}^3$, along which the differentiation occurs. The term $(\R^\mathrm{T} \mathbf{e})$ on the right-hand side can be factored out of the integral.

The rotational gradient function $df_\mathrm{SC}/d\R : \mathrm{SE}(3) \rightarrow \mathds{C}^3$ is more difficult to formulate, since $\mathrm{SO}(3)$ is not a vector space and cannot be globally parameterized by a single continuous 3D grid. To obtain a local parametrization, the tangent direction at $\R \in \mathrm{SO}(3)$ is obtained as $\R \Omega$ where $\Omega \in \mathfrak{so}(3)$ can be represented by a skew-symmetric matrix $[\Omega]_{3\times3}$, and $\mathfrak{so}(3)$ denotes the Lie algebra, which is a vector space tangent to $\mathrm{SO}(3)$ at the identity rotation. Without getting into much detail, we present the rotational gradient as
\begin{equation}
    \langle \frac{d f_\mathrm{SC}}{d \R}, \mathbf{e} \rangle = - \int_{\mathds{R}^3} \rho_1(\mathbf{p}) \Big[ \nabla \rho_2 \circ \R^\mathrm{T} (\mathbf{p} - \mathbf{t}) \Big] \cdot \Omega^\ast (\mathbf{p} - \mathbf{t}) d\volume, \label{eq_5c}
\end{equation}
where $\mathbf{e} = \R \mathbf{u}$ and $\mathbf{u} \in \mathds{R}^3$ is the dual vector of $\Omega \in \mathfrak{so}(3)$, and $\Omega^\ast = \R \Omega \R^\mathrm{T}$ is called the action of $\R^\mathrm{T}$ on $\Omega$.
The affinity gradient $\nabla \rho_2 = \nabla \rho_\sigma(\mathbf{p}; S_2)$ used in the integrands of both (\ref{eq_5b}) and (\ref{eq_5c}) is computed from (\ref{eq_3a}).

The 3D translational and rotational gradient vectors can be computed in a componentwise fashion by substituting for the base vectors $\mathbf{e} \in \{\mathbf{e}_1, \mathbf{e}_2, \mathbf{e}_3 \}$ one at a time in (\ref{eq_5b}) and (\ref{eq_5c}).
The complete 6D gradient $\nabla f_\mathrm{SC} : \mathrm{SE}(3) \rightarrow \mathds{C}^6$ is defined as $\nabla f_\mathrm{SC} = (df_\mathrm{SC} / d\R, df_\mathrm{SC}/ d\mathbf{t})$.

\subsection{Geometric Energies}

The described generic and continuous score distribution over the configuration space $\mathrm{SE}(3)$ rewards shape complementarity and penalizes collision and separation. This enables defining an artificial potential energy function $E_\mathrm{G} \propto \Re\{f_\mathrm{SC}\}$ for use in real-time haptic assembly.

\paragraph{Energy Function.}

We define the `geometric energy' function $E_\mathrm{G}: \mathrm{SE}(3) \rightarrow \mathds{R}$ simply as:
\begin{equation}
    E_\mathrm{G}\big( \motion; S_1, S_2 \big) = -\gamma_\mathrm{SC} \cdot \Re \Big\{ f_\mathrm{SC}\big( \motion; S_1, S_2 \big) \Big\}, \label{eq_6}
\end{equation}
where $\Re\{\cdot\}$ stands for the real-part, and the constant $\gamma_\mathrm{SC} > 0$ is provided to scale the dimensionless score function $f_\mathrm{SC}$ to proper energy units, before applying it to objects of certain mass and inertia properties in a scene, bearing in mind the possibility of other forces being present.

One could appreciate an interesting analogy between this artificial, purely geometric energy field, and physical energy fields such as the electrostatic effect. It immediately follows that the product of affinity functions can be conceptualized as a complex `geometric potential,' which applies on a complex `geometric charge' density, whose magnitude is dictated by the $\lambda-$function in (\ref{eq_2}). Using this analogy, on the one hand, when charges on the two objects are imaginary numbers of the same sign, there is a positive-real contribution to the energy, implying a {\it repulsion} force (Fig. \ref{figure2} (b)). On the other hand, when the charges are imaginary numbers of opposite signs, they contribute a negative-real energy, which indicates an {\it attraction} force (Fig. \ref{figure2} (c)). Both attractive and repulsive effects decay with distance, due to the inverse-square law embedded in the $\phi-$kernel in (\ref{eq_2}).

\paragraph{Force and Torque.}
The conservative `geometric force' and `geometric torque' are obtained as the gradient of the energy function with respect to the translational and rotational motion, respectively:
\begin{align}
    \mathbf{F}_\mathrm{G}\big( (\R, \mathbf{t}); S_1, S_2 \big) &= -\frac{d E_\mathrm{G}}{d \mathbf{t}} = +\gamma_\mathrm{SC} \Re \left\{ \frac{d f_\mathrm{SC}}{d \mathbf{t}} \right\}, \label{eq_9a} \\
    \mathbf{T}_\mathrm{G}\big( (\R, \mathbf{t}); S_1, S_2 \big) &= -\frac{d E_\mathrm{G}}{d \R} = +\gamma_\mathrm{SC} \Re \left\{ \frac{d f_\mathrm{SC}}{d \R} \right\}. \label{eq_9b}
\end{align}
This can be consolidated into the 6D general force/torque vector $(\mathbf{T}_\mathrm{G}, \mathbf{F}_\mathrm{G}) = - \nabla E_\mathrm{G}$, where $\nabla E_\mathrm{G} : \mathrm{SE}^3 \rightarrow \mathds{R}^6$ is the complete 6D energy gradient defined as $\nabla E_\mathrm{G} = -\gamma_\mathrm{SC} \Re \{ \nabla f_\mathrm{SC}\}$.

\section{Implementation}

In this section we present an implementation of our method for a particularly simple representation (namely, triangular mesh B-reps), along with the underlying algorithms and data structures. We briefly overview the computational complexity of each step, and refer the reader to \cite{Behandish2014} for more details.

\subsection{Representation} \label{sec_rep}

As described in Section \ref{sec_form}, our method is independent of the representation scheme used to describe the solid objects in the scene. Any representation scheme that satisfies the informational completeness requirement \cite{Requicha1980a} can be used, provided that it supports the means to 1) compute unsigned Euclidean distance queries to the boundary points; and 2) a PMC test \cite{Tilove1980a} to correct the sign of the distance function---both to an adequate accuracy with respect to the smallest surface features. For the sake of simplicity, we present the implementation for B-reps in the form of triangular mesh surfaces with oriented boundary elements.
In particular, we use a data structure that contains the following:

\begin{itemize}
    \item[\textbullet] {\it combinatorial structure:} the adjacency and orientation information for the boundary faces, edges, and vertices (e.g., using oriented half-edge or barycentric decomposition data structures); and
    \item[\textbullet] {\it metric information:} the coordinates of the boundary vertices and (optionally) vertex normals, which can be used to obtain the face normals defining a consistent orientation for the boundary manifold.
\end{itemize}

Given a solid $S \in \rset$, we denote the underlying space\footnote{The `underlying space' of a cell complex is the union of all cells in that complex. For a 2D meshed surface embedded in 3D, it means the 2D subspace of the $3-$space occupied by all faces, edges, and vertices of the triangulation \cite{Requicha1977a}.}
of a triangulation that approximates its boundary $\partial S$ with $\Delta_n(S) = \bigcup_{j = 1}^n \delta_j$, where the closed triangles are denoted by $\delta_j~(1 \leq j \leq n)$, the number of triangles (i.e., faces) is $n$, and the number of edges and vertices are both $O(n)$ \cite{Requicha1977a}. We use the NETGEN library \cite{Schoberl1997} to triangulate the boundary of solid parts exported in STEP format from any commercial CAD software. The boundary vertices need to be sampled with adequate density to capture the local geometric features of the shape.

\subsection{Preprocessing}

For a single query point $\mathbf{p} \in \mathds{R}^3$, the sequence of computations is 1) unsigned distance queries from the $n$ triangles; 2) the PMC test to correct the distance sign for $\zeta-$mapping in (\ref{eq_1}); and 3) applying the $\phi-$kernel in (\ref{eq_2}) followed by the integration in (\ref{eq_3}). The same sequence needs to be carried out over a 3D uniform grid $G_m(S)$ of $m$ nodes.
Let $\mathbf{q}_j \in \Delta_n(S) ~ (1 \leq j \leq n)$ denote a point (e.g., the mid-point) on a triangle $\delta_j \subset \Delta_n(S)$, whose unit normal is $\mathbf{n}_j \in \mathds{R}^3$ and surface area is $\delta \area_j > 0$. Let $\mathbf{p}_i \in G_m(S) ~ (1 \leq i \leq m)$ denote a arbitrary grid node. Then (\ref{eq_3}) can be approximated by the following discrete form:
\begin{equation}
    \rho_\sigma(\mathbf{p}_i; S) \approx \sum_{j = 1}^n \phi_\sigma \Big[ \xi_i + \ii \eta_{i,j} \Big] \cos \theta_{i,j} \delta \area_j, \label{eq_10}
\end{equation}
where $\xi_i = \xi(\mathbf{p}_i; S)$, $\eta_{i,j} = \eta(\mathbf{p}_i, \mathbf{q}_j)$, and $\cos \theta_{i,j} = \mathbf{v}_{i,j} \cdot \mathbf{n}_j$, in which $\mathbf{v}_{i,j} = (\mathbf{p}_i - \mathbf{q}_j)/\eta_{i,j}$.
A similar discrete from can be obtained for the affinity gradient integral in (\ref{eq_3a}). It is important to note that the approximation in (\ref{eq_10}) is reliable if the spatial angle by which the triangle $\delta_j$ is observed from $\mathbf{p}$ is small (i.e., $\cos \theta_{i,j} \delta \area_j \ll \eta_{i,j}^2$) which is not necessarily true for grid nodes that are close to the $\partial S$ surface, constituting only a small fraction of all grid nodes. For those points, the triangle can be recursively subdivided into smaller faces, until an upperbound criterion on the spatial angle of observation is reached. Assuming that the number of recursions is $O(1)$, computing (\ref{eq_10}) takes $O(n)$ basic steps. Therefore, the computation of the SDF over the entire grid $G_m(S)$ takes $O(mn)$ steps.
The grid cell size should be small enough to capture the geometric features of the shape by the SDF, which implies a lowerbound on $m$. As described in \cite{Behandish2014}, this can be sped up to $O(m'n)$ where $m' \ll m$ by using adaptively sampled query points, for instance over an octree $Q_{m'}(S)$ composed of $m'$ nodes but with the same minimum cell size as that of $G_m(S)$.

For distance computations, we use Havoc3D \cite{Hoff1999}, which cumulatively computes the unsigned distance field $|\xi_i| ~ (1 \leq i \leq m)$ using interpolation-based polygon rasterization on the graphics hardware via OpenGL depth-buffer. For sign determination (i.e., PMC) we used the method in \cite{Klein2009} which is based on winding numbers defined in terms of signed spatial angles. We implemented both PMC and SDF field computations in parallel on the CPU, assigning different chunks of $G_m(S)$ to different processors, using OpenMP.\footnote{We report on a similar implementation on the Graphics Processing Unit (GPU) in \cite{Behandish2015a}.}
The SDF field needs to be precomputed offline, {\it only once} per each rigid part or subassembly, hence can be done with high precision with little concern about the computation time. 

\begin{figure*}
    \centering
    \includegraphics[width=\textwidth]{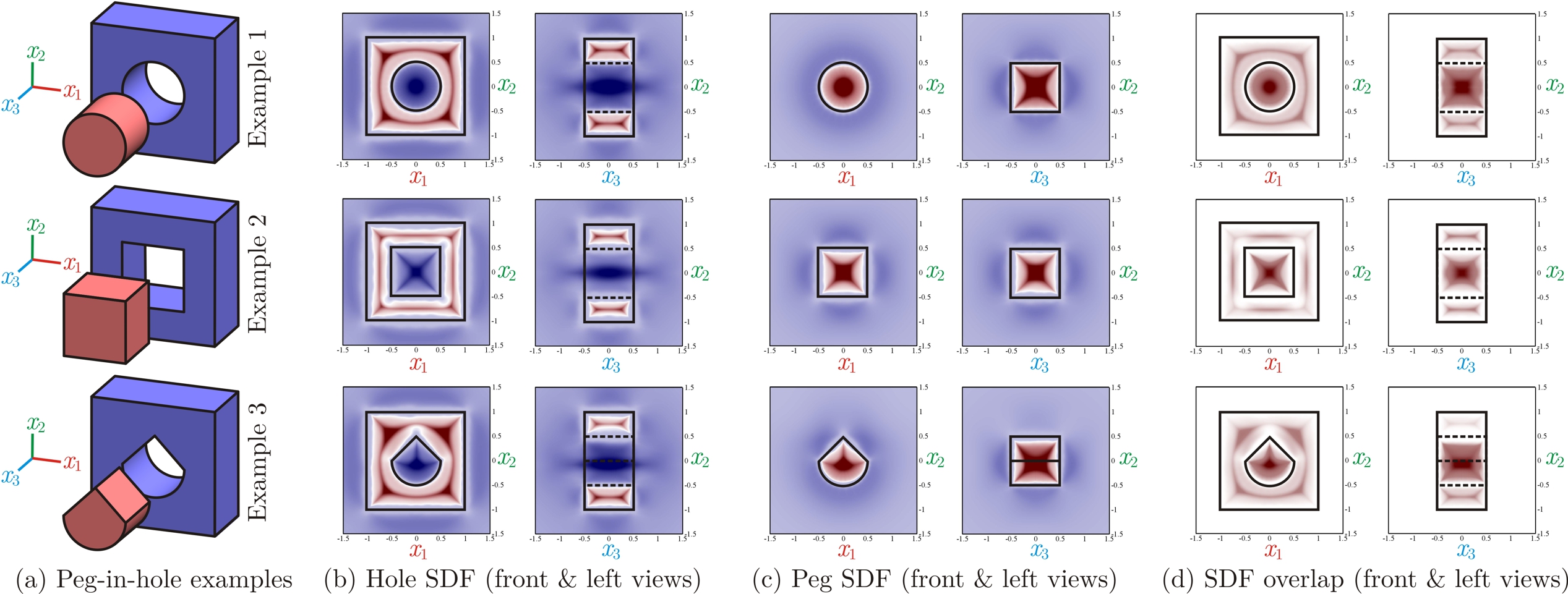}
    \caption{Three peg-in-hole assemblies (a), their SDFs (imaginary-parts) (b, c), and their spatial overlap (real-part) (d).} \label{figure4}
\end{figure*}

\subsection{Cross-Correlation}

Let the two assembly partners $S_1$ and $S_2$ be represented with triangular meshes $\Delta_{n_1}(S_1)$ and $\Delta_{n_2}(S_2)$ composed of $n_1$ and $n_2$ triangles, respectively. Assuming that the SDF fields for the individual objects are precomputed separately over $G_{m_1}(S_1)$ and $G_{m_2}(S_2)$ grids attached to each body, at every instance of the dynamic simulation with $\motion_1, \motion_2 \in \mathrm{SE}(3)$ the score integral in (\ref{eq_5_0}) can be discretized into
\begin{equation}
    f_\mathrm{SC}(\motion_{1,2}; S_{1,2}) \approx \sum_{i = 1}^{m} \rho(\motion_1^{-1} \mathbf{p}_i; S_1) \rho(\motion_2^{-1}\mathbf{p}_i; S_2) \delta V, \label{eq_11}
\end{equation}
where $\mathbf{p}_i \in G_m(\motion_1 S_1 \cap \motion_2 S_2)$ is a node on a grid sampled uniformly over the intersection of the moved objects, and $\delta V = vol(G_m) / m$ is the cell volume of this grid. The SDFs are interpolated from the precomputed values in (\ref{eq_10}). To save in interpolation time, the integration grid is picked as a subset of the smaller SDF grid, hence $m \leq \min\{ m_1, m_2 \}$ and computing (\ref{eq_11}) takes $O(m)$ basic steps.
The score gradient in (\ref{eq_5b}) and (\ref{eq_5c}), needed for the guidance forces and torques in (\ref{eq_9a}) and (\ref{eq_9b}), can be discretized in a similar fashion. Alternatively, one could approximate the gradient using the finite difference method (FDM) by multiple computations of (\ref{eq_11}) after applying small translational and rotational variations, along each of the 3 coordinate axes one at a time, to the relative transformation $\motion = \motion_1^{-1} \motion_2 \in \mathrm{SE}(3)$.

We parallelize the force and torque computations on the CPU using OpenMP. Although the performance scales almost linearly with the number of cores, the running times are not adequately small to keep up with the $1$ kHz haptic rendering loop. The simplest solution is to precompute the geometric energy $E_\mathrm{G}$ and/or forces and torques $(\mathbf{T}_\mathrm{G}, \mathbf{F}_\mathrm{G})$ over a sampling of relative transformations in $\mathrm{SE}(3)$, and interpolate the sample in real-time. This is not practical (both in terms of time and memory) for a 6D configuration space, even with the powerful computers available today.
Fortunately, for most assembly scenarios the motion during the insertion phase is constrained to 1 or 2 DOF. For example, if the rotational space is limited to a finite number of permissible relative orientations, the field can be precomputed and stored over a 3D translational sampling in $\mathrm{T}(3)$ for each orientation (i.e., a section through the configuration space). However, this approach goes against the philosophy of avoiding the multiphase approach and manual specifications, from which we set off to pursue this method. We recently presented an alternative implementation in \cite{Behandish2015a,Behandish2015b} that uses GPU-accelerated fast Fourier transforms (FFT) to enable real-time computations, the discussion of which is beyond the scope of this paper.

\section{Results \& Discussion} \label{sec_results}

We first demonstrate the effectiveness of the proposed approach for simple classical peg-in-hole examples. Figure \ref{figure4} (a) shows the three examples made of cylindrical pegs of circular, rectangular, and combined cross-sections, which were tested in assembly against their complementary holes. The geometric fit in all three cases is exact (i.e., zero clearance).

\begin{figure*}
    \centering
    \includegraphics[width=\textwidth]{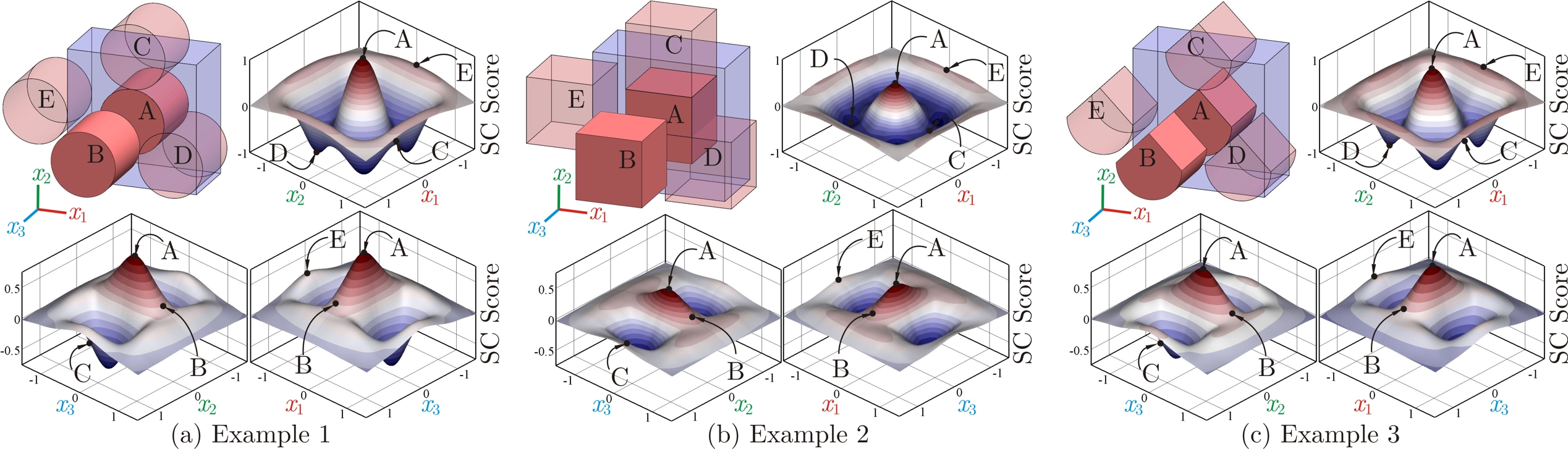}
    \caption{The shape complementarity score variations versus biaxial relative translation of the peg with respect to the hole.} \label{figure5}
\end{figure*}
%

\subsection{Skeletal Overlaps}

In Fig. \ref{figure4} (b, c), the individual SDF maps of the parts are plotted only for their imaginary-parts. As expected, each part has the highest positive-imaginary SDF values at the proximity of the high-prong internal MA branches (the red spots) and the highest negative-imaginary SDF values at the proximity of the high-prong external MA branches (the blue spots). The latter are weaker in intensity due to the choice of $\mathfrak{p} = \lambda_2/ \lambda_1 = 3.0$ in (\ref{eq_2}). It can be verified that the complementary features have similar SDF distributions on one part's interior and the other part's exterior, resulting in the `hot spots' on the geometric energy density map given in Fig. \ref{figure4} (d), which dominate the `dark spots.'

\paragraph{Geometric Guidance.}
In the simplest case of Example 1, as the user brings the peg closer to the opening of the hole to perform the assembly task, the geometric force-field attracts the peg into the hole and tries to align the high-density SDF regions, i.e., enforce coaxiallity of the two cylindrical faces. The circular symmetry of the cross-section results in a circular symmetry of the SDF, hence the force-field imposes no orientation preference around the axis of the hole, resulting in a partially constrained motion that resembles a cylindrical joint. However, in the case of Example 2, the cross-shaped skeletal form creates an additional orientation preference; hence the force-field tries to align the four corners of the two complementary objects. This results in a partially constrained motion that resembles a prismatic joint. In the case of Example 3, the shape descriptors appear as a combination of the two cases, with part of the geometry being indifferent to rotations around the hole, while another feature contributes energy terms to align the sharp corner.

It is clear from the above discussion that our SDF descriptors serve as generic replacements for the abstract virtual fixtures \cite{Rosenberg1993} (e.g., cylinder axes in Example 1 and diagonal planes in Example 2), variations of which were previously implemented in \cite{Tching2010a} for haptic assembly guidance, and were limited to simple geometric constructs. The skeletal branches formed automatically in our development serve as abstractions of the functional surfaces \cite{Iacob2011} for arbitrarily complex shapes, in contrast to the {ad hoc} characterizations in \cite{Iacob2008}. As illustrated by Example 3, combinations of guiding mechanisms naturally appear with no theoretical limitation on the complexity of the assembly features. Furthermore, the force-field incorporates both collision response (as a repulsive force in the case of interpenetration) and assembly assistance (as an attractive force in the vicinity of the hole) in a single model that enforces geometric constraints.
Hence the hybrid approach based on two separate phases for free motion and precise insertion \cite{Vance2011} is integrated into a single model, eliminating the need to switch between the two using `blending' algorithms.

\paragraph{Energy Variations.}
Figure \ref{figure5} plots the shape complementarity score variations (only the real-parts) due to the translational motion of the peg relative to the hole along the 3 Cartesian axes. To enable visual illustration, the motion in each plot is restricted to a plane (i.e., changing only 2 out of 3 position coordinates at a time, of the peg with respect to the hole). It is clear that the score is maximal at configuration A (i.e., the zero translation, where the best fit occurs according to our visual judgment), as expected from the definition. Other configurations are also sketched on the score profile, such as axis-aligned removal of the peg at B (resulting in a decay of score from A to B along the $x_3-$axis), collision at C and D, and contact (but little shape complementarity) at E.

\begin{figure*}
    \centering
    \includegraphics[width=\textwidth]{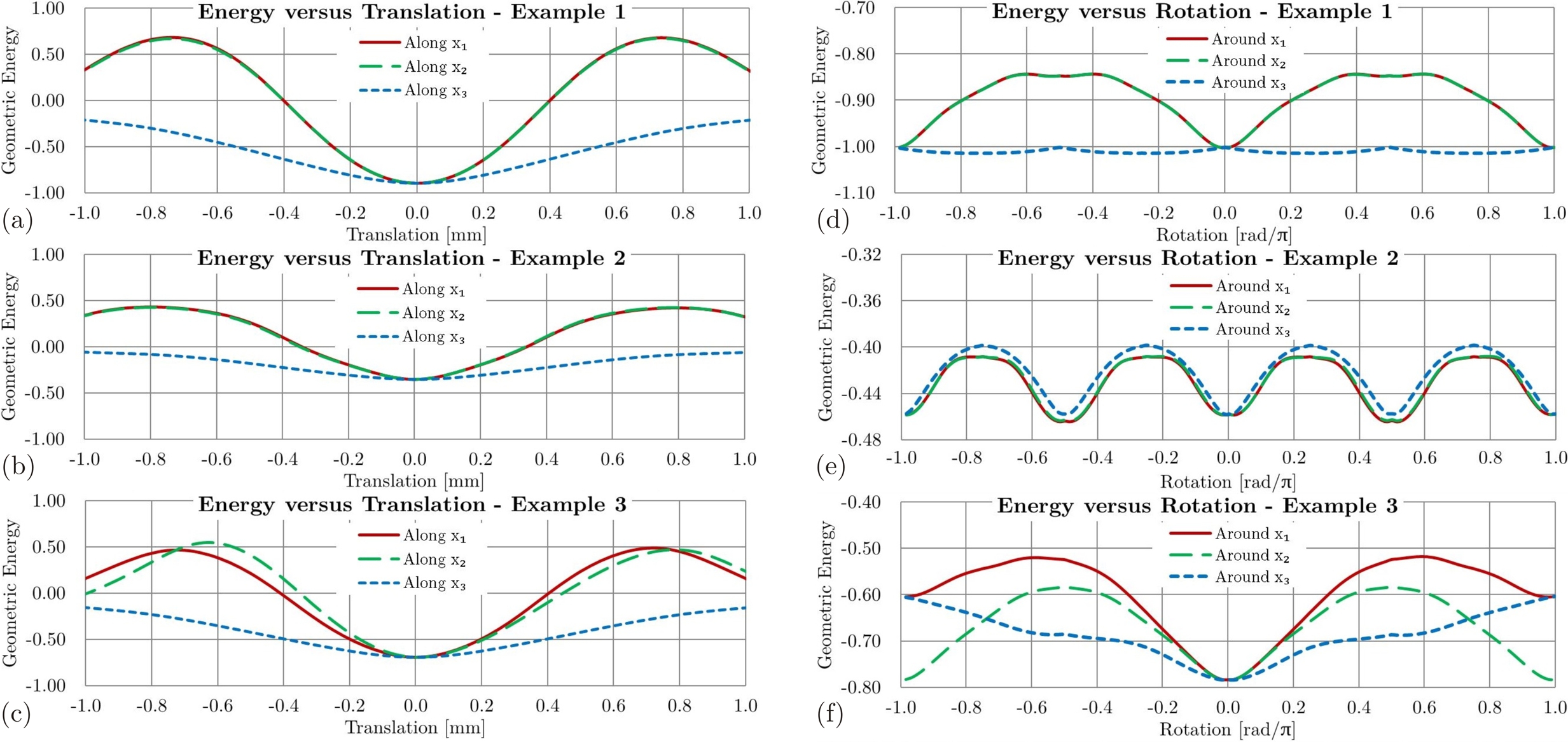}
    \caption{The geometric energy variations versus uniaxial relative translation and rotation of the peg with respect to the hole.} \label{figure6}
\end{figure*}

Figure \ref{figure6} (a) shows the corresponding geometric energy variations due to the same translational motions, this time moving the peg along one Cartesian axis at a time. These correspond to sections through the 2D plots in Fig. \ref{figure5}, except with a signed coefficient due to the definition in (\ref{eq_6}). Figure \ref{figure6} (b) shows a similar set of geometric energy variation diagrams, plotted versus the rotational motions around one Cartesian axis at a time. For both translational and rotational motions, there is an evident equivalence between $x_1-$ and $x_2-$axes in Examples 1 and 2, as expected from the symmetrical shapes, which is not the case for Example 3 due to its different geometry. One can also notice the indifference of the circular cross-section to rotations around the axis of the hole in Example 1, and four equivalent configurations for the cubic peg with $90^\circ$ phase difference in Example 2.

\paragraph{Constraint Stiffness.}
It is interesting to note that the size of the convex region of the geometric energy profile around the local minimum (characterizing the equilibrium point) can be conceptualized as the {\it diameter} of the geometric constraints, i.e., the degree of proximity necessary for the constraint to become activated for insertion. The second-order differential properties of the energy profile are directly related to the practical {\it stiffness} of the constraint enforcement in the VE. For instance, a Taylor series expansion of the energy function over the translational space $\mathrm{T}(3)$ (i.e., for fixed rotation) around the stable equilibrium configuration $\mathbf{t}_0 \in \mathrm{T}(3)$ has the form $E_\mathrm{G} = E_{\mathrm{G}, \min} + (\mathbf{t} - \mathbf{t}_0) \cdot H(\mathbf{t}_0) (\mathbf{t} - \mathbf{t}_0) + O(\| \mathbf{t} - \mathbf{t}_0 \|_2^3)$, noting that $d E_\mathrm{G} / d\mathbf{t} (\mathbf{t}_0) = \mathbf{0}$ where the Hessian matrix $[H(\mathbf{t}_0)]_{3\times3}$ carries the stiffness elements (i.e., tensile/compressive resistance in the diagonal elements, and shear resistance in the off-diagonal elements).\footnote{Note than no constraint can be rigidly and strictly satisfied due to the electromechanical restrictions at the haptic device level. Even the DOF-limiting equality constraints are typically enforced by rendering resistance forces using a spring/damper model that penalizes the violation, whose stiffness is upperbounded due to the servo-loop rate of $1$ kHz \cite{Perret2013}.}
A similar expansion is possible over the tangent space $\mathfrak{so}(3)$ to $\mathrm{SO}(3)$ to obtain the rotational stiffness matrix.
Both the diameter and stiffness can be adjusted by a proper setting of the thickness factor $\sigma > 0$ and coefficients $\lambda_{1,2}$ in (\ref{eq_2}).

\subsection{Haptic Experiments}

Finally, we conduct a few simple experiments to {\it feel} the applicability of the technique in real haptic-assisted assembly applications. Two simple experiments are carried out on a SensAble$^\circledR$ PHANTOM$^\circledR$ Omni$^\circledR$ device, namely:

\begin{enumerate}
    \item {\it Collision Test}: the test is conducted by simply pushing the peg against the walls of the hole in random directions, in an attempt to disturb it from the proper fit configuration and penetrate into the walls of the object with the hole. The user tries to do this with complete control and steadiness, approximately once every two seconds. 
    \item {\it Snap Test}: the user carelessly moves the peg around the entrance of the hole, with random and uncontrolled (but gentle) impacts with the end-effector, positioning the peg in proximity of the proper fit configuration approximately once every second. The force field is expected to react immediately and snap the peg into the proper position.
\end{enumerate}

\begin{figure*}
    \centering
    \includegraphics[width=\textwidth]{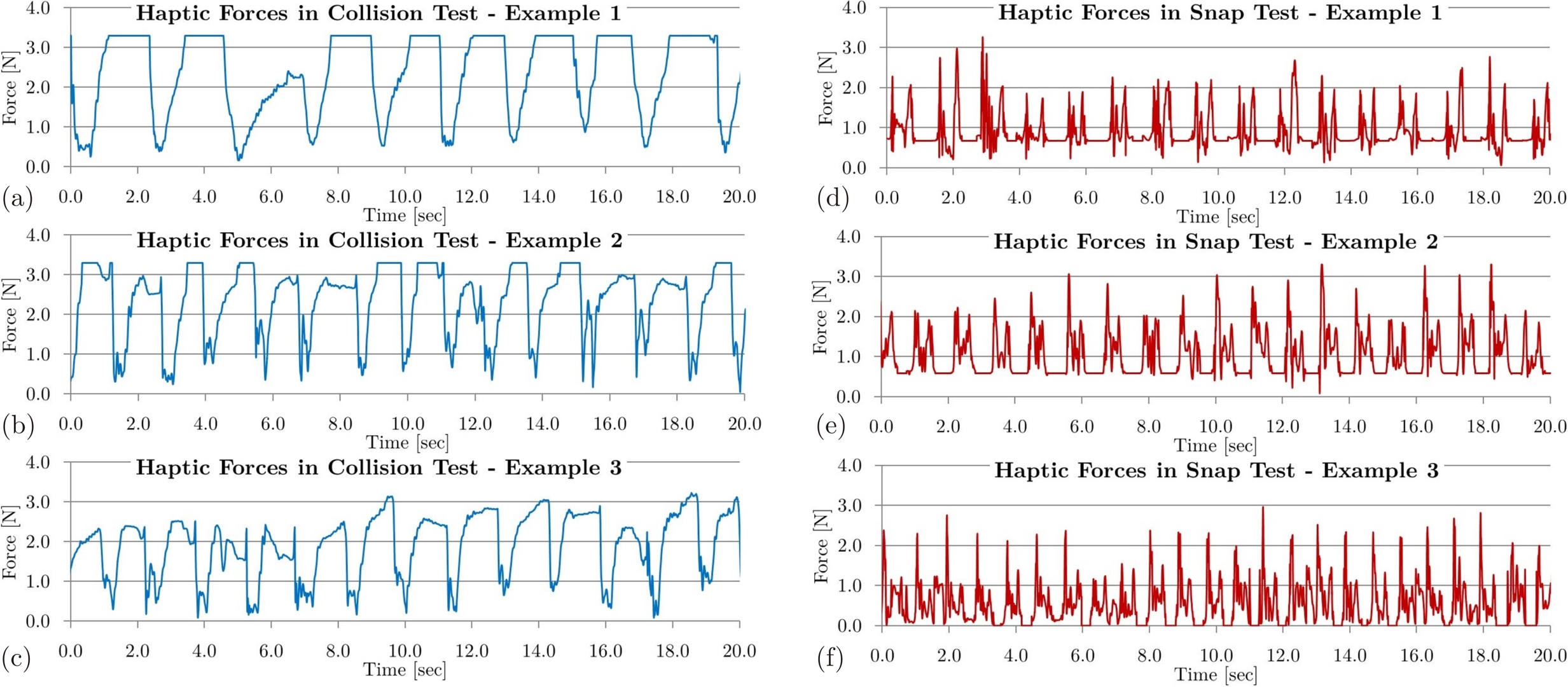}
    \caption{Haptic force feedback versus time, for collision test (a--c) and snap test (d--f) for peg-in-hole examples in Fig. \ref{figure4}.} \label{figure7}
\end{figure*}

The tests are carried out only in 3 translational DOF, and the experimentation with rotational motions requires to be addressed in future studies with a 6 DOF haptic device. The force magnitude versus time is plotted in Fig. \ref{figure7} (a) for the first experiment. The performance is satisfactory, with accurate geometric alignment up to observable precision, smooth and continuous repulsive force feedback resisting penetration in all directions, and a smaller attractive force resisting the peg leaving the hole along the axis. The results of the second experiment are plotted in Fig. \ref{figure7} (b). In this case, the response was effective in snapping the object into position with very rare occasions of undesirable vibration or `buzzing.' The haptic servo-loop rate was maintained within the acceptable range of $1.0 \pm 0.3$ kHz at all times.

\section{Conclusion}

Lately, the dominant direction in haptic assembly has been aligned with a hybrid approach, separating the simulation into a `free motion' phase, using unilateral (i.e., inequality) `physical constraints' originated from collision detection; and a `fine insertion' phase, using bilateral (i.e., equality) `geometric constraints' introduced artificially to limit the DOF. While the former fail to produce dynamically stable guidance for low-clearance insertion, the latter are either dependent on {a priori} manual specifications by the user, or are limited to simple algebraic subspaces (e.g., planar, cylindrical, spherical, or conical) that can be identified automatically from CAD semantics using heuristic algorithms.
We proposed a novel paradigm that unifies the two modes into a single interaction, by expanding the collision penalty function to a generic `geometric energy' field that not only penalizes the configurations with interpenetration (hence produces the repulsive collision response), but also rewards the configurations with high {\it shape complementarity} (hence produces the attractive guidance forces). We accomplished this by formulating the energy function as a cross-correlation of new descriptors of shape, to which we referred as the SDF. The SDF interactions can be conceptualized as generic replacements for {ad hoc} virtual fixtures or simplistic mating constraints, and apply to objects of arbitrary shape.
We showed that this approach automatically ensures a continuous transition between collision response in free movement to insertion guidance in low-clearance or precise-fit assembly, avoiding the two-phase approach along with its several drawbacks---including the failure to prevent collision events outside the insertion site, and the need for blending the force feedback during the switch.

The unified paradigm provides a promising alternative direction for solving virtual assembly problems in general (and for haptic rendering, in particular). It opens up new research opportunities to investigate faster implementations and verify effectiveness for 6 DOF manipulation in complex and crowded assembly scenes.

\section{Acknowledgment}

This work was supported in part by the National Science Foundation grants CMMI-1200089, CMMI-0927105, and CNS-0927105.
The responsibility for any errors and omissions lies solely with the authors.
The identification of any commercial product or trade name does not imply endorsement or recommendation.

\bibliographystyle{asmems4}
{\scriptsize \bibliography{CDL-TR-15-08}}

\end{document}